# A single-molecule approach to ZnO defect studies: single photons and single defects

N. R. Jungwirth, Y. Y. Pai, H. S. Chang, E. R. MacQuarrie, and G. D. Fuchs

Cornell University, Ithaca, New York 14853, USA

**Abstract:**

Investigations that probe defects one at a time offer a unique opportunity to observe properties and dynamics that are washed out of ensemble measurements.  Here we present confocal fluorescence measurements of individual defects in Al-doped ZnO nanoparticles and undoped ZnO sputtered films that are excited with sub-bandgap energy light.  Photon correlation measurements yield both antibunching and bunching, indicative of single-photon emission from isolated defects that possess a metastable shelving state.  The single-photon emission is in the range ~560 – 720 nm and typically exhibits two broad spectral peaks separated by ~150 meV.  The excited state lifetimes range from 1 – 13 ns, consistent with the finite-size and surface effects of nanoparticles and small grains.  We also observe discrete jumps in the fluorescence intensity between a bright state and a dark state.  The dwell times in each state are exponentially distributed and the average dwell time in the bright (dark) state does (may) depend on the power of the exciting laser.  Taken together, our measurements demonstrate the utility of a single-molecule approach to semiconductor defect studies and highlight ZnO as a potential host material for single-defect based applications.

## I. Introduction and Background

### A. Motivation

ZnO is a material of choice for a wide range of applications in optoelectronics [1–3] due to its piezoelectricity, wide bandgap (~3.3 eV), and large exciton binding energy. Because defects often play a pivotal role in determining the bulk electrical, optical, and magnetic properties of materials, the identification and function of defects in ZnO remains the subject of intense study [4–14]. For ZnO to have its anticipated impact, long-standing questions regarding the defects, for example their role in unintentional n-type conductivity [4,8,9], must be resolved.

Prior ZnO defect studies have examined photoluminescence [15–27], electron paramagnetic resonance [12,18,20,28–32], and magnetic resonance [12,19,23,30] of defect *ensembles*. Such measurements integrate over many instances of a particular defect species and are consequently unable to observe defect-to-defect variability and single-defect dynamics. Minor differences in the emission spectrum of individual defects, which result from sample inhomogeneities, may produce an ensemble spectrum that appears broadened or structureless. Ensemble measurements also risk convolving signal from defects of different structural origin, making it difficult to assign a particular bulk property to a particular defect species.

In this work we evade the ensemble average by studying individual ZnO defects, one at a time, using single-molecule microscopy techniques [33–41]. We observe excited state lifetimes ranging from 1 - 13 ns across doped and undoped ZnO samples.  We also observe discrete jumps in the fluorescence intensity between a bright state and a dark state.  The dwell times in the bright (dark) state are exponentially distributed with an average dwell time that does (may) depend on laser power.  Our results are unobtainable by ensemble methods and demonstrate that single-molecule microscopy techniques are a useful tool for probing properties that are hidden in more conventional studies.

This work is also motivated by the potential application of defects in quantum information technology [42–46] and precision sensing [47–50].  Single defects in wide bandgap semiconductors constitute a spatially localized quantum system that may be individually addressed.  The nitrogen-vacancy (NV) center in diamond, a point defect that consists of a substitutional nitrogen atom adjacent to a lattice vacancy, is one example.  NV centers possess a spin triplet ground and excited state and a metastable singlet state, all of which are contained inside the diamond bandgap [51,52].  The spin state of individual NV centers may be initialized, measured, and controlled at room temperature [53], making them promising qubit candidates.  Although NV centers have proven to be a valuable spin system for fundamental investigations of quantum information science, the diamond host material is one constraint to the scalability of NV-based applications.  Single-crystal diamond substrates are small, expensive, and have minimal epitaxial compatibility with other materials.  Moreover, standard nanophotonic fabrication techniques are not easily extended to diamond substrates.  These limitations are less pertinent for a material like ZnO, which is inexpensive, abundant, and has well-developed fabrication and growth techniques [54,55]. These practical considerations have spurred a search for NV-like defects in more conventional wide bandgap semiconductors [46] most notably ZnO [56] and SiC [57–59]. In the case of ZnO, Morfa et. al [56] recently reported antibunching, bunching, and spectral measurements from ZnO defects that they ascribed to a Zn vacancy. Despite this promising start, a more thorough understanding of ZnO single-defect properties, including photodynamics and coupling with the local environment, is required before ZnO defects can be seriously regarded as a candidate semiconductor defect-based qubit.

### B. Isolating Single Defects

The bandgap energy, $E_G$, of a semiconductor corresponds to the theoretical minimum photon energy that may be absorbed by that material. In real materials this absorption threshold is typically reduced by point defects, such as a lattice vacancy, that introduce spatially localized states inside the bandgap [60–62]. In this work [Fig. 1(a)], we excite defect fluorescence by focusing sub-$E_G$ green light (2.3 eV) onto ZnO ($E_G$ = 3.3 eV) using a numerical aperture 0.7 microscope objective. The green light selectively excites defect states because these 2.3 eV photons do not possess sufficient energy to excite carriers across the ZnO bandgap. The subsequent fluorescence, which is red-shifted from its absorption energy due to phonon coupling, is detected using either a single-photon sensitive avalanche photodiode (APD) or a spectrometer. This approach is sufficiently sensitive to measure in the single-defect limit.

To ensure we study defects one at a time, we measure the second-order correlation function, $g^{(2)}(\tau)$, of the emitted photons. For stationary processes this function is given by [63]

$$g^{(2)}(\tau) = \frac{p_e(\tau|0)}{p_e^{st}}, \tag{1}$$

where $p_e(\tau|0)$ is the conditional probability of emitting a photon at time $t = \tau$ supposing a photon was emitted at time $t = 0$ and $p_e^{st} = p_e(\pm\infty|0)$ is the steady state probability of emitting a photon. Because a defect returns to its ground state when emitting a photon, the limiting behavior for a single defect is $p_e(0|0) = 0$. Thus, an isolated defect has $g^{(2)}(0) = 0$ and $g^{(2)}(\pm\infty) = 1$. The dip at $\tau = 0$, known as an antibunching dip, is the signature of single-photon emission; it results from the fact that once a quantum system emits a photon, it cannot emit a second until it has been re-excited and subsequently re-relaxes to the ground state.

Fig. 1(a) illustrates how $g^{(2)}(\tau)$ is measured using a Hanbury Brown and Twiss setup. The fluorescence passes through a 50/50 beam splitter and is detected by one of two APDs. The delay between successive photon detection events at each APD is computed by a time correlated single photon counting (TCSPC) module. The normalized distribution of these delay times is equivalent to $g^{(2)}(\tau)$ in the limit that $\tau$ is much smaller than the average time between detection events [64]. Fig. 2 shows several representative $g^{(2)}(\tau)$ curves produced using this method. Note that $g^{(2)}(\pm\infty) = 1$ as expected but $g^{(2)}(0) \neq 0$. Experimental non-idealities including detector dark counts, background counts, detector jitter, and finite histogram bin width prevent most experimental antibunching dips from reaching their theoretical minimum value of zero. Because two identical single-photon emitters within the detection volume of the microscope theoretically yield $g^{(2)}(0) = 0.5$ [65], dips extending below $0.5$ are typically regarded as verification that only one quantum emitter is being studied. However, defect-to-defect variations in fluorescence intensity and the existence of blinking effects (discussed later) allow small ensembles of defects to produce antibunching dips extending below $0.5$. We note that all data presented in this work exhibit antibunching dips extending below 0.4.

### C. Rate Model

It is useful to associate a ground state $|1\rangle$, an excited state $|2\rangle$, and a metastable shelving state $|3\rangle$ with a given defect. These states are not necessarily single-electron states and they may possess spin degrees of freedom. Three possible state configurations are illustrated in Fig. 1(b). In one case [Fig. 1(b), middle] the states are completely contained inside the bandgap; we shall refer to this as an "internal transitions". In separate scenarios [Fig. 1(b), left and right], some of the states are positioned in either the conduction or valence band; we shall refer to these as "band transitions". In all three configurations, a defect in the ground state $|1\rangle$ may absorb a photon of sufficient energy and enter the excited state $|2\rangle$. The excited state offers two decay pathways: a relaxation directly to the ground state or a non-radiative intersystem crossing via the metastable state to the ground state. Photon emission may accompany the direct $|2\rangle \to |1\rangle$ relaxation with a likelihood known as the quantum

efficiency. Neglecting coherences, a rate model describes state population evolution [66]:

$$\frac{d}{dt}\begin{pmatrix} P_{|1\rangle} \\ P_{|2\rangle} \\ P_{|3\rangle} \end{pmatrix} = \begin{pmatrix} -r_{12} & r_{21} & r_{31} \\ r_{12} & -r_{21} - r_{23} & 0 \\ 0 & r_{23} & -r_{31} \end{pmatrix} \begin{pmatrix} P_{|1\rangle} \\ P_{|2\rangle} \\ P_{|3\rangle} \end{pmatrix}. \quad (2)$$

Here $P_{|i\rangle}$ is the probability of occupation of state $|i\rangle$ and $r_{ij}$ is the transition rate from state $|i\rangle$ to state $|j\rangle$. Note that while the transition rates may vary during the course of an experiment due to local fluctuations, for this analysis we ignore the time dependence of the transition rates.

Eq. 2 may be used to model the second-order correlation function, $g^{(2)}(\tau)$, introduced earlier (Eq. 1). Because the probability of emitting a photon is proportional to $P_{|2\rangle}$, and an emission event is accompanied by a relaxation to the ground state, $p_e(\tau|0)$ is proportional to $P_{|2\rangle}(\tau)$ with initial condition $(P_{|1\rangle}, P_{|2\rangle}, P_{|3\rangle})(t=0) = (1,0,0)$ [66]. We find that in the absence of background counts

$$g^{(2)}(\tau) = 1 + Ae^{\lambda_1 \tau} - (1+A)e^{\lambda_2 \tau}, \quad (3)$$

where $A$ is determined by the transition rates and $\lambda_1, \lambda_2$ are the nonzero eigenvalues of the rate matrix appearing in Eq. 2. Note that because Eq. 3 is bi-exponential, photon bunching is possible. Photon bunching, evident in the uppermost curve of Fig. 2(b), occurs when $g^{(2)}(\tau) > 1$. Bunching is absent in 2-state systems because such systems only permit mono-exponential solutions for $P_{|2\rangle}(\tau)$. Thus, the observation of photon bunching confirms that the system being studied has at least three states.

The transition rates appearing in Eq. 2 and illustrated in Fig. 1(b) characterize the defect's optical dynamics. They determine the defect's fluorescence intensity and provide insight into its oscillator strength. The excited state lifetime, which is defined as $\tau_{|2\rangle} = (r_{21} + r_{23})^{-1}$, is of particular importance and may be determined directly using a 532 nm pulsed excitation and time-correlated photon counting. A representative measurement for a single defect is shown in Fig. 1(c). The solid line represents an exponential fit to the data. The inset, which displays the same data on a log scale, verifies that the decay is indeed a pure exponential and is characterized by a single time constant, $\tau_{|2\rangle}$.

**D. Photodynamics**

Single-molecule microscopy requires temporal averaging because of the comparably low photon signal that is detected from one defect (~30 kCounts/s). Consequently, properties that vary over a timescale shorter than a measurement will remain hidden. Some temporal variations may be monitored in real time, however, by observing the fluorescence intensity, or photon emission rate, under continuous excitation. In principle, a defect's fluorescence intensity is completely determined by the transition rates and follows a Poisson distribution. However, single-photon sources, including the ZnO defects investigated here, rarely exhibit stable Poissonian emission. Instabilities in fluorescence intensity resulting from spectral diffusion [67–69], changes in the local environment [69,70], photobleaching [69,71], and the presence of a long-lived metastable state [70,72–74] have been reported, and sophisticated models have been developed to address these issues [38,63,66,75–78]. Although fluorescence intensity instabilities have been viewed negatively in the context of single-photon sources, they may potentially be exploited to understand the system under study or as sensing channel [79]. For example, a defect's fluorescence intensity may serve as a sensitive environmental probe of its nanoscopic environment. Alternatively, a defect's fluorescence intensity may potentially be controlled via magnetic fields, electric fields, or stress.

Here we examine blinking, whereby the fluorescence intensity exhibits discrete jumps between a bright state and a dark state. Blinking is a common feature of isolated quantum systems [80,81] and has been reported in many systems including quantum dots [82–85], single molecules [69,70,72,74,86–90], nanocrystals [71,91], defects in SiC [59], and diamond NV centers [79,92,93]. The mechanism(s) responsible for blinking may be investigated by constructing the distribution of dwell times spent in the bright state, $\tau_{BS}$, and the dark state, $\tau_{DS}$, for several laser powers or temperatures. Previous investigations have separately reported a power law [70,71,82,85,88] and exponential [73,79,83,89] distribution for blinking dwell times in quantum dots, nanocrystals, single molecules, and single defects. Because a quantum system emits no photons while trapped in a metastable state, a long-lived metastable state ($\tau_{|3\rangle} > \mu s$) is a commonly cited cause for blinking [69,73,74,86,89]. In this case one expects the dwell times in each state to be exponentially distributed and the average dwell times, $\langle \tau_{BS/DS} \rangle$, to obey [69]

$$\frac{1}{\langle \tau_{BS} \rangle} = \frac{r_{12} r_{23}}{r_{21} + r_{23}}, \tag{4}$$

$$\frac{1}{\langle \tau_{DS} \rangle} = r_{31}, \tag{5}$$

where it is assumed that $r_{12} \ll r_{21} + r_{23}$. Alternatively, jumps from a bright state to a dark state could result from the emergence of non-radiative $|2\rangle \to |1\rangle$ decay channels that serve to reduce $\tau_{|2\rangle}$ and quench the fluorescence intensity by reducing the quantum efficiency [71,94–96]. Because in this scenario the fluorescence intensity is quenched but not extinguished, the emitter is always optically active, and one expects single-photon emission in both the bright state and the dark state (in contrast to blinking caused by a long-lived metastable state, which only exhibits antibunching in the bright state). As a third possibility, blinking may result from either a thermal- or photo- induced ionization that couples two distinct charge states of distinguishable fluorescence intensity [93]. Note that we have not presented an exhaustive or mutually exclusive list of blinking mechanisms and separate mechanisms may coexist.

## II. Experiment

### A. Sample Details

We investigated both nanoparticle (NP) ZnO and sputtered ZnO films. The NP samples were purchased from Sigma Aldrich and include undoped and 6% Al-doped by weight. The NPs were suspended in methanol prior to being drop-cast onto a fused silica substrate. We observe a lower density and fluorescence intensity of defects in undoped ZnO NPs compared with their doped counterparts. Though we observed antibunching in each NP type, we did not observe $g^{(2)}(0) < 0.5$ in the undoped NP samples. Thus all NP data presented result from Al-doped NPs. X-ray diffraction (XRD) measurements indicate that the Al-doped NPs are wurtzite structure and have a mean size of 75 nm.

ZnO films were grown on fused silica substrates by reactive DC sputtering. The sputter chamber was pumped to a base pressure of $5 \cdot 10^{-7}$ Torr. Subsequently Ar and O were introduced into the chamber with partial pressures of 4 mTorr and 1 mTorr, respectively. The film thickness was monitored using a quartz crystal monitor that was calibrated using a contact profilmeter. XRD measurements indicate that our sputtered films are wurtzite, polycrystalline with a mean grain size of 50 nm, and are ⟨002⟩ orientated.

All presented data (NPs and sputtered films) are from samples annealed at 500° C in air for 30 minutes. We also studied unannealed samples, but we were unable to isolate any single defects prior to annealing. The fused silica substrates used for both sputtered film growth and NP drop-casting were chosen for their low background fluorescence intensity and absence of native single-photon emitters.

The emission spectrum of individual ZnO defects were investigated and several of these spectra are shown in Fig. 3. Each spectrum has been normalized and the dashed line at ~580 nm denotes the cutoff of a long-wave pass filter edge. Ten of the 16 spectra we investigated are well characterized by two Gaussian-distributed peaks separated by roughly 150 meV. Each of these spectral peaks originates from the same defect, not from another nearby defect or from background emission. The results are included in Table I. We found the location of the low (high) energy peak to be in the range of 1.82 – 1.97 eV (1.91-2.08 eV). These findings are consistent with prior work [56], although we note that our results are of a higher energy than those reported previously. We did not observe a room-temperature zero phonon line, which is consistent with the strong defect-phonon coupling present in the spectral emission [97].

**B. Transition Rates**

Although the excited state lifetime may be directly determined using pulsed excitation, that approach does not measure the metastable state lifetime ($1/r_{31}$) and cannot decouple the direct ($r_{21}$) and intersystem crossing ($r_{23}$) pathways for leaving the excited state. These considerations motivate us to determine the transition rates by measuring $g^{(2)}(\tau)$ for several different laser powers. The excitation rate $r_{12}$ is proportional to laser power and may therefore be readily varied over several orders of magnitude. We simultaneously fit a series of $g^{(2)}(\tau)$ (antibunching) curves, each of which was acquired from the same defect but at a different laser power, using Eq. 3. In the fit $r_{21}$, $r_{23}$, and $r_{31}$ are kept constant across all curves and $r_{12}$ is constrained to increase linearly with excitation power. The best-fit for the series of curves was determined using a chi-squared minimization procedure. We emphasize that the fit is not obtained by directly modifying the eigenvalues $\lambda_1$ and $\lambda_2$. Instead, we use the analytic form of these eigenvalues and vary the transition rates as free parameters, thus enabling determination of all transition rates simultaneously.

Two examples of an antibunching power series are displayed in Fig. 2. In each curve the antibunching dip at $\tau = 0$ extends below the dashed line located at 0.5, verifying that a single defect is being probed. The data from Fig. 2(a) were acquired from the same defect for which we measured the excited state lifetime using pulsed excitation [Fig. 1(c)]. By fitting this antibunching power series, we determine an excited state lifetime of 4.8 ± 0.2 ns. This result is in close agreement with the pulsed excitation measurement of 4.5 ± 0.3 ns, thus corroborating the efficacy of our approach.

Several noteworthy trends are displayed in Fig. 2(b). Firstly, the sharpness of the antibunching dip increases with increasing excitation power. Additionally, photon bunching, which often emerges as excitation power is increased, is evident and confirms that the defect has at least three states. Note that though the data in Fig. 2(a) show no signs of bunching, there may nonetheless still be three states present. That is, the absence of bunching in $g^{(2)}(\tau)$ neither verifies nor refutes that a third state exists.

Using the method described above we determined the transition rates for 19 individual ZnO defects. Our results are displayed in Table I. We computed uncertainties in lifetime measurements by using the chi-squared curvature matrix as outlined by Bevington [98] and neglect other experimental sources of error. Of the defects presented, 14 are from NP samples and five are from sputtered samples. To our knowledge, this is the first report of single-photon emission from sputtered ZnO, which is promising for quantum applications because sputtered films are readily engineered and may be integrated for device applications. Seven of the 14 NP-based defects studied displayed signs of bunching and one of the five sputtered film defects exhibited bunching. While our analysis allows us to determine the metastable state lifetime, the best-fit is not highly sensitive to this parameter, and thus we are much less certain about the metastable lifetime compared to the excited state lifetime. In cases where bunching is apparent but not pronounced enough to enable a reliable determination of the metastable state lifetime, we have listed the metastable state lifetime as unknown. The metastable state lifetimes reported here are in the range 40 to 300 ns which is in agreement with those reported previously [56].

The excited state lifetime, which we observed to range from 1.7 ns to 13 ns, exhibits much defect-to-defect variability. The variability in the rate of leaving the excited state, $1/\tau_{|2\rangle}$, is illustrated in Fig. 4. The individual rates for each defect investigated are represented as markers situated below the histogram bars. Note that for a conduction band transition [Fig. 1(b), left] we expect $1/\tau_{|2\rangle} = n \cdot C_n$, where $n$ is the local carrier density and $C_n$ is the defect's electron capture coefficient. Though we have not determined the carrier concentration in our samples, in particular the NPs, we anticipate the carrier density of our undoped sputtered films to be less than that of the heavily Al-doped NPs. For the limited doping values investigated, Fig. 4 does not indicate a strong correlation between doping concentration and $1/\tau_{|2\rangle}$. This suggests that we are studying defects with an excited state that is decoupled from the conduction band. However, reports of high unintentional n-type conductivity ($n > 10^{19} cm^{-3}$) in as-grown, undoped ZnO films [99,100] indicate that the carrier concentration in ZnO is highly sensitive to the particular growth details, making a direct comparison between NPs and sputtered films problematic. Moreover, several studies indicate that in addition to affecting the carrier density, Al dopants in ZnO may induce structural changes, shift $E_G$, or alter the equilibrium defect concentrations [101–106]. Consequently, a comparison between undoped and heavily Al-doped samples, even when grown by identical methods, is nontrivial. Thus, further doping dependent studies are necessary to confirm that the defects studied are indeed decoupled from the conduction band. Nonetheless, we tentatively attribute the wide spread in measured $\tau_{|2\rangle}$ to variations in NP/grain size, geometry, surface chemistry, and to differences in the defect to surface distance rather than variations in the local carrier density. It has been shown for NV centers in diamond that as the diamond host is reduced in size, non-radiative relaxation channels, which are not present in bulk diamond, emerge that serve to reduce the NV excited state lifetime [107]. Additionally, a nanodiamond's specific geometry may affect the excited state lifetime of its NV centers [108]. In fact, there is evidence that the radiative decay rates, non-radiative decay rates, and quantum efficiencies of NV centers are especially variable in nanodiamonds as large as $100\ nm$ [109]. Because we are studying nanoparticles (~75 nm) and polycrystalline films (~50 nm grains), these effects may play a substantial role in governing the wide spread of $\tau_{|2\rangle}$ values that we observe.

### C. Photodynamics (Blinking)

Figs. 5(a)-(c) depict temporal variations in the fluorescence intensity of three distinct ZnO defects. In Figs. 5(a) and (b) (defects 3 and 4, respectively) the fluorescence intensity exhibits discrete jumps between a bright state and a dark state. Each state is well defined and we observe Poissonian photon shot-noise statistics within each state. This situation of bright/dark blinking is contrasted by Fig. 5(c). Here the fluorescence follows a more complicated trajectory characterized by at least four levels. We have characterized the fluorescence of each defect in Table I as "stable", meaning purely Poissonian statistics, "on/off", meaning blinking between a bright and a dark state that each exhibit Poissonian statistics, and "fluctuates", meaning the fluctuations in fluorescence intensity exceed the Poissonian statistical expectation. In the following analysis we focus exclusively on bright/dark blinkers, such as those seen in Figs. 5(a) and (b), and characterize the distribution of dwell times spent in each state before a blinking event occurs.

Blinking between a bright state and a dark state suggests a scenario whereby an individual defect is jumping between an "on" state where it emits single photons and an "off" state where it does not emit any photons at all. To test if this is indeed the case, we measured $g^{(2)}(\tau)$ separately for the bright state and the dark state by post-filtering our photon arrival times. The results of this state-selective measurement are displayed in Figs. 5 (d) and (e). Fig. 5(d), which was solely acquired while defect 3 was in the bright state, shows clear antibunching and confirms that the bright state corresponds to single-photon emission. Fig. 5(e), which was solely acquired while defect 3 was in the dark state, shows no evidence of antibunching and indicates that the dark state corresponds to a relatively small flux of uncorrelated background photons. Thus when defect 3 is in the dark state, it is either not fluorescent or is fluorescent but emits into a spectral window not detectable by our setup. In either case, single-photon emission statistics are associated only with the bright state and the terms on/off and bright/dark blinking may be used interchangeably.

We investigated the distribution of dwell times for both the bright state ($\tau_{BS}$) and dark state ($\tau_{DS}$) of 2 separate bright/dark blinking defects. For each time bin the defect state was determined by comparing the number of photons detected to a threshold value. This approach requires careful selection of binning times and thresholds because improper selection of these values may introduce artifacts into the measured dwell times [78,110,111]. Here we find that the dwell times in both the bright and dark states are exponentially distributed [Figs. 6(a) and (b), respectively] with a time constant that corresponds to the average dwell time of each state.

We measured the average rate of leaving the bright state ($\langle\tau_{BS}\rangle^{-1}$) and the dark state ($\langle\tau_{DS}\rangle^{-1}$) for defect 5 for several laser powers [Figs. 6(c) and (d), respectively]. Fig. 6(c) indicates that $\langle\tau_{BS}\rangle^{-1}$ increases as laser power is increased whereas Fig. 6(d) shows no correlation between $\langle\tau_{DS}\rangle^{-1}$ and laser power. These trends are consistent with defect 5 blinking due to a long-lived metastable state ($\tau_{|3\rangle} \sim 50$ ms) (Eqs. 4 and 5). In this scenario, each blinking event from the bright state to the dark state corresponds to the collapse of the defect's quantum state to the metastable state $\left(BS \xrightarrow{|2\rangle \rightarrow |3\rangle} DS\right)$. Alternatively, these trends are also consistent with a photo- or thermal- induced band transition that couples the excited state $|2\rangle$ to the conduction band. In such a scenario, each blinking event from the bright state to the dark state corresponds to an ionization event. Incidentally, if ionization is the mechanism, then the bright state is the energetically favorable charge state because the $BS \rightarrow DS$ transition requires optical excitation whereas the $DS \rightarrow BS$ transition does not. The data presented is insufficient to uniquely determine the cause of blinking for defect 5. Temperature dependent blinking studies may aid in identifying the active blinking mechanism.

We performed an identical dwell time analysis for defect 3, which exhibits considerably longer dwell times than defect 5.  As before, the dwell times in each state are exponentially distributed and $\langle \tau_{BS} \rangle^{-1}$ increases with increasing excitation power [Fig. 6(e)].  However, we find that $\langle \tau_{DS} \rangle^{-1}$ also increases with increasing laser power [Fig. 6(f)] for defect 3.  This behavior indicates that defect 3 does not blink due to a long-lived metastable state.  Additionally, because both $\langle \tau_{BS} \rangle^{-1}$ and $\langle \tau_{DS} \rangle^{-1}$ approach zero as laser power is reduced, a purely band-assisted mechanism is unlikely.  One blinking mechanism consistent with Figs. 6(e) and (f) is a charge transfer between defect 3 and a neighboring acceptor or surface trap.  In this scenario the $BS \rightarrow DS$ transition corresponds to an electron escaping the defect and becoming localized at a nearby site.  The ejected electron then serves to reduce the local carrier density via Coulombic repulsion and enhance the dark state lifetime.

While the active blinking mechanisms for defect 3 and defect 5 have not been determined, their qualitative differences suggest that multiple flavors of dark state may exist [86,89,94].  If multiple dark state varieties exist, then a single defect may potentially couple to multiple types.  The dark state dwell times for defect 5 [Figs. 6(b) and (d)] are exponentially distributed with $\langle \tau_{DS} \rangle \approx 50 \; ms$ for all laser powers studied.  Thus observing any dwell time greater that ~10 s should be exceedingly improbable because only ~$10^{-85}$% of all observed dwell times are expected to be greater than 10 s. Despite this, dark state dwell times in excess of one minute were observed for defect 5. One such anomalous dwell time is denoted by the arrow in Fig. 7(a). This figure shows the fluorescence intensity of defect 5 over a time span of 500 s. Several anomalous dwell times are apparent in addition to the one that has been singled out. Note that the dense regions of the graph, such as that occurring between approximately 0 s and 150 s, are regions of rapid blinking between the bright and dark states. This is illustrated in Fig. 7(b), which displays a sub-second time window of Fig. 7(a). Evidently defect 5 is strongly coupled to a short-lived dark state and is weakly coupled to a distinct, anomalously long-lived dark state. The anomalous dark state dwell times evident in Fig. 7(a) occur too infrequently to be studied statistically. Nonetheless, the observation of such anomalous dwell times is compelling evidence that several distinct mechanisms may jointly describe the rich blinking dynamics of a single defect.

### III. Conclusion

We have characterized the room temperature properties and dynamics of single defects in Al-doped ZnO NPs and undoped sputtered ZnO films. The defects investigated exhibit a range of behavior with some commonalities and some differences. Similarities include absorption of 532 nm light, single-photon emission in the red, excited state lifetimes in the range ~ 1 - 13ns, and metastable state lifetimes in the range ~ 40 – 300 ns. Differences include variations in the locations of peaks in the emission spectrum, the values of the transition rates, and defect photodynamics. For defects with identical or similar structure, one may anticipate less variation in the lifetimes and emission spectra than we report here. However, it has been shown for nanodiamond NV centers that the host crystal's size and geometry greatly influence the radiative decay rate, non-radiative decay rate, and quantum efficiency [107–109]. Because we are studying ZnO nanoparticles and polycrystalline grains with a range of sizes and surface geometries, similar mechanisms could explain the wide range of excited state lifetimes observed here. Additionally, variations in the local environment, such as the presence of hydrogen interstitials, have been predicted to alter the emission spectra of Zn vacancies in ZnO [112]. Thus it is possible that the defects investigated here have identical or similar structure despite the wide range of observed behavior. Recent theoretical advancements in first-principles calculations of defect emission [113] are especially promising for understanding the defect-to-defect variability that we observe.

Though our observed excited state lifetimes range from ~ 1 - 13 ns, we discern no correlation between doping concentration and excited state lifetime. This would suggest that we study defects with an excited state that is largely decoupled from the conduction band. However, because our heavily doped NPs were produced differently than the undoped sputtered films, and because high unintentional n-type conductivity ($n > 10^{19} cm^{-3}$) in undoped films has been reported [99,100], a conclusion based on the data presented is still premature and is the subject of ongoing study. Nevertheless, we note that prior theoretical work on defects in ZnO has focused on band transitions that couple distinct defect charge states [4–10]. These calculations shed light on the electrical properties of ZnO but do not address the electronic excitations of a particular defect charge state (internal transitions). If we indeed study defects with internal transitions rather than band transitions, then a thorough understanding of ZnO may require a more detailed knowledge of internal defect transitions than presently exists. Lacking a first-principles prediction for ZnO defects matching our observations, we follow the assignment of Morfa et. al [56] that the defects studied are Zn vacancy related because Zn vacancies have been shown to exhibit red emission [12].

We also report cases of discrete blinking between a bright state, where the defect is optically active and emits single photons, and a dark state, where only a relatively small flux of background photons are incident on the detector. Whereas the rate of blinking from the bright state to the dark state increases with laser power, the rate of blinking from the dark state to the bright state may or may not increase with laser power. These two distinct dark state behaviors likely result from distinct blinking mechanisms, which is supported by statistical evidence that a single defect may couple to multiple dark states. Future temperature dependent experiments may offer valuable clues regarding the nature of blinking in ZnO single defects.

This work illustrates the utility of a single-molecule approach to semiconductor defect studies. We focus on individual defects, thus obtaining results that are unattainable in an ensemble study. In contrast, an ensemble photoluminescence measurement of the same defects studied here would yield a weighted sum of the individual defect spectra that would be featureless except for a broad "red" photoluminescence band. Moreover, the variability in defect transition rates and the blinking dynamics would be averaged out. Our observations also suggest that the properties of single defects in ZnO are very sensitive to the local environment and that single-defect studies could be useful for understanding local variations in ZnO.

## Acknowledgements

We gratefully acknowledge insightful conversations with John Lyons, Daniel Steiauf, Luke Gordon, Audrius Alkauskas, Anderson Janotti, and Chris Van de Walle from the University of California at Santa Barbara. This work was supported by the Cornell Center for Materials Research with funding from the NSF MRSEC program (DMR-1120296), by the National Science Foundation (DMR-1254530), and by the Department of Energy Office of Science Graduate Fellowship Program (DOE SCGF), made possible in part by the American Recovery and Reinvestment Act of 2009, administered by ORISE-ORAU under contract no. DE-AC05-06OR23100.

**Table I:**

| Defect | Sample | $\tau_{|2\rangle}$ (ns) | $\tau_{|3\rangle}$ (ns) | Spectral Peak Locations (eV) | Spectral Peak Separation (meV) | Fluorescence Intensity |
|---|---|---|---|---|---|---|
| 1 | NP | 4.8 +/- 0.2 | | (1.91 , 2.04) | 130 | Stable |
| 2 | NP | 3.2 +/- 0.1 | 52 +/- 5 | | | Fluctuates |
| 3 | NP | 4.53 +/- 0.03 | | | | On/Off |
| 4 | NP | 2.17 +/- 0.03 | 280 +/- 200 | (1.88 , 2.06) | 180 | On/Off |
| 5 | NP | 1.66 +/- 0.03 | | (1.94 , 2.08) | 340 | On/Off |
| 6 | NP | 1.29 +/- 0.05 | | (1.94, 2.07) | 130 | On/Off |
| 7 | NP | 3.37 +/- 0.04 | | | | On/Off |
| 8 | NP | 1.92 +/- 0.03 | Unknown | | | Stable |
| 9 | NP | 1.79 +/- 0.04 | 100 +/- 15 | (1.93 , 2.08) | 150 | Stable |
| 10 | NP | 1.66 +/- 0.04 | 44 +/- 2 | | | Fluctuates |
| 11 | NP | 4.6 +/- 0.2 | 100 +/- 20 | (1.97 , 2.11) | 140 | Fluctuates |
| 12 | NP | 3.25 +/- 0.03 | 221 +/- 5 | (1.89 , 2.03) | 140 | Fluctuates |
| 13 | NP | 3.6 +/- 0.3 | | | | Fluctuates |
| 14 | NP | 7.3 +/-0.2 | | (1.89 , 2.06) | 170 | Fluctuates |
| 15 | 20 nm SP | 3.3 +/- 0.1 | | (1.95 , 2.08) | 130 | On/Off |
| 16 | 71 nm SP | 5.2 +/- 0.3 | | | | Fluctuates |
| 17 | 71 nm SP | 13.4 +/- 0.6 | | | | Fluctuates |
| 18 | 71 nm SP | 8.8 +/- 0.2 | Unknown | | | Fluctuates |
| 19 | 71 nm SP | 6.9 +/- 0.03 | | (1.82 , 1.91) | 90 | Fluctuates |

Summary of the 19 single defects studied. The sample types include nanoparticle (NP) ZnO and sputtered ZnO films (SP) for which the film thicknesses have been listed in $nm$. All lifetimes were determined by simultaneously fitting a power series of $g^{(2)}(\tau)$ curves and all uncertainties are calculated using the chi-squared curvature matrix [98]. When bunching was evident but a reliable determination of $\tau_{|3\rangle}$ was not possible, we have listed $\tau_{|3\rangle}$ as "Unknown". For defects whose emission spectra is characterized by two broad Gaussian peaks, we have listed the midpoint of each peak in eV and have also included the peak-to-peak separation in meV. Lastly, the fluorescence intensity of each defect is characterized as "Stable", meaning well described by Poissonian statistics, "On/Off" meaning discrete blinking between two "Stable" states of distinguishable fluorescence intensity, and "Fluctuates", meaning intensity fluctuations that cannot be described as "Stable" or "On/Off".

**Figure 1:**

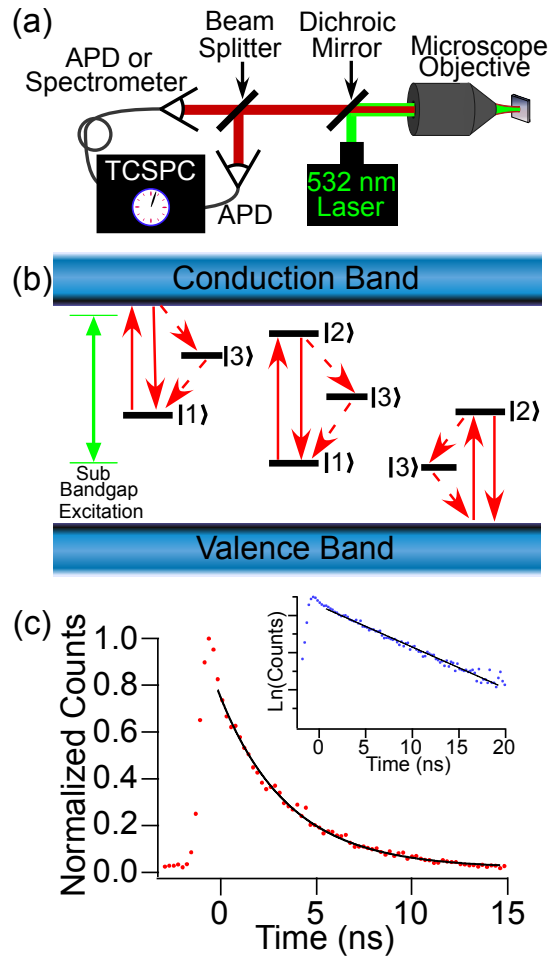

(a) Schematic of confocal microscope used to probe single defects. Sub-bandgap energy green light excites local defect states and the subsequent emission is detected by an APD sensitive to single photons or by a spectrometer. A time correlated single photon counting (TCSPC) module enables measurement of $g^{(2)}(\tau)$ for verification that a single defect is being studied. (b) Three models of defect levels consistent with the antibunching/bunching measurements we present. Each model admits a ground state $|1\rangle$, an excited state $|2\rangle$, and a metastable state $|3\rangle$. The allowed transitions are illustrated as arrows and the dashed arrows correspond to non-radiative decay pathways. In the middle model, all defect states are contained inside the bandgap and thus none of the transitions couple distinct defect charge states. In the left (right) model, the excited state $|2\rangle$ is above (below) the conduction (valence) band minimum (maximum) and thus the $|1\rangle \rightarrow |2\rangle$ excitation involves an ionization of the defect. (c) Lifetime measurement of defect 1 using a pulsed laser. The inset is the same data on a log scale and serves to verify that the decay is a pure exponential.

**Figure 2:**

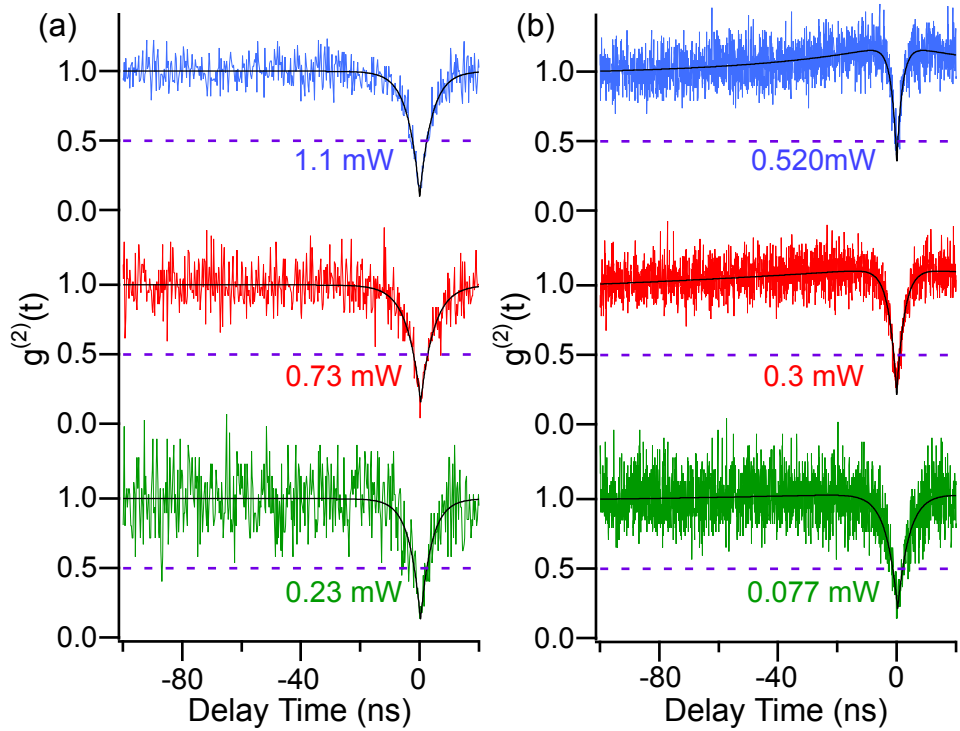

Representative $g^{(2)}(\tau)$ curves for defect 1 (a) and defect 2 (b) acquired for several different laser powers. In each curve the antibunching dip at $\tau = 0$ extends below the dashed line at 0.5, verifying that a single defect is being studied. Note in (b) that as laser power is increased the antibunching dip sharpens and bunching $\left(g^{(2)}(\tau) > 1\right)$ becomes evident, verifying at least 3 defect states exist.

**Figure 3:**

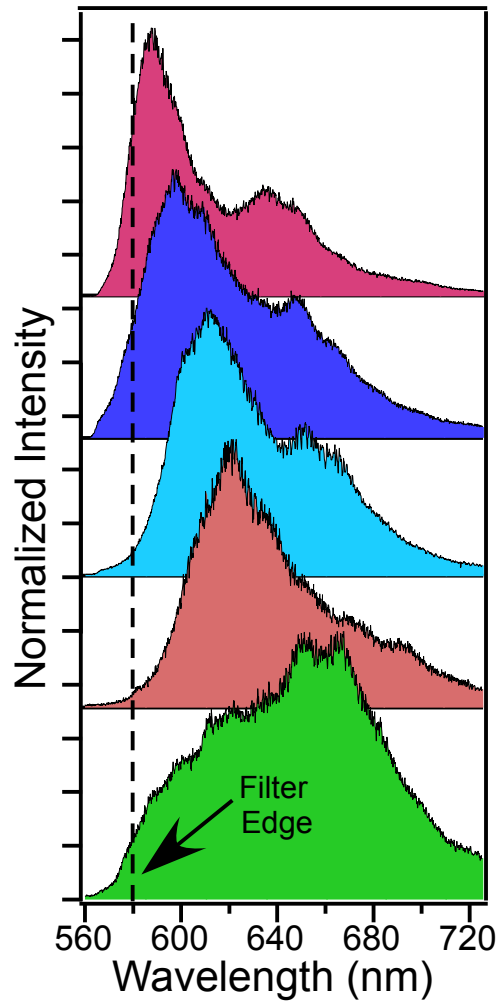

Normalized emission spectra for 5 of the 19 individual defects investigated. The emission is plotted in the range ~ 560 - 720 nm and is typically characterized by two broad peaks that may be fit by Gaussian functions. Each spectrum arises exclusively from a single defect and not two or more. The dashed line at ~ 580 nm denotes the cutoff of a long-wave pass filter edge

**Figure 4:**

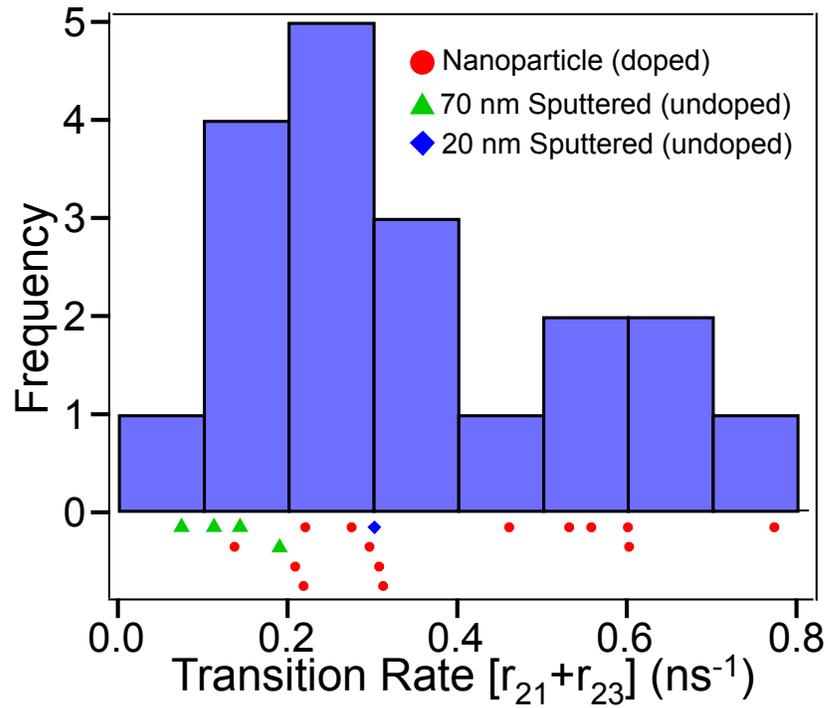

Histogram of observed transition rates ($1/\tau_{|2\rangle}$). Below the histogram bars, the transition rate for each defect studied has been represented by a marker. Note the significant overlap in rates for defects in undoped sputtered films (triangles and diamonds) and 6% Al-doped NPs (circles). We tentatively attribute the wide spread in transition rates to variations in grain/NP size, geometry, distance to the surface, and surface chemistry rather than variations in the local carrier density.

**Figure 5:**

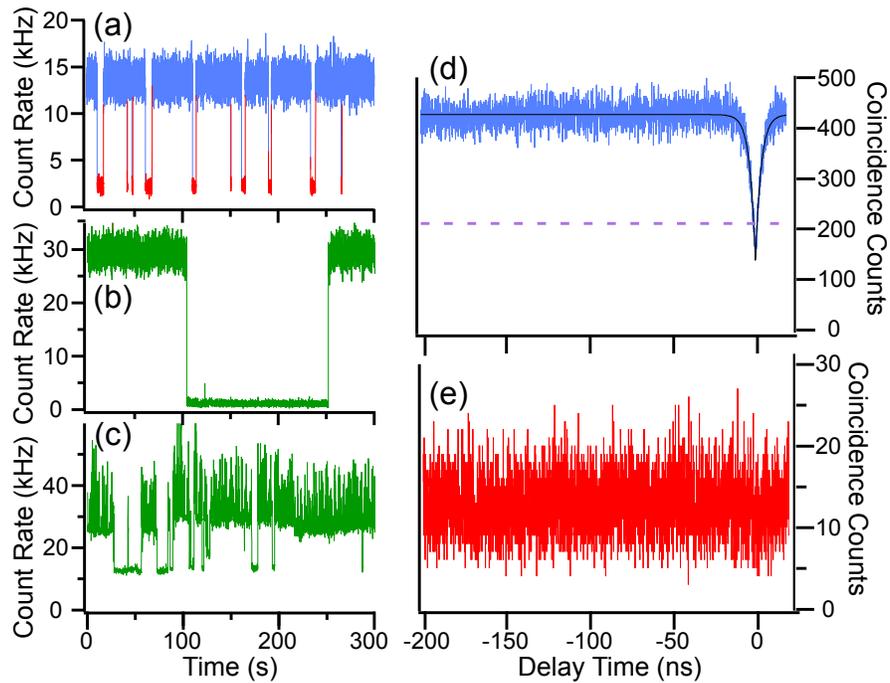

(a-c) Examples of temporal dynamics in the fluorescence intensity for three distinct defects over a time span of 300 s. Defect 3 (a) and defect 4 (b) each exhibit discrete jumps, or blinking, between a bright state and a dark state whereas the fluorescence intensity seen in (c) follows a much more complicated trajectory. For defect 3 (a), we measured $g^{(2)}(\tau)$ separately for the bright state (d) and the dark state (e). The presence of antibunching in (d) verifies that the defect is optically active and emits single photons while in the bright state. The absence of antibunching in (e) indicates that the defect is inactive, and not merely quenched, while in the dark state. Note the differences in counts between (d) and (e), indicating that the vast majority of coincidence counts are registered while the defect is in the bright state.

**Figure 6:**

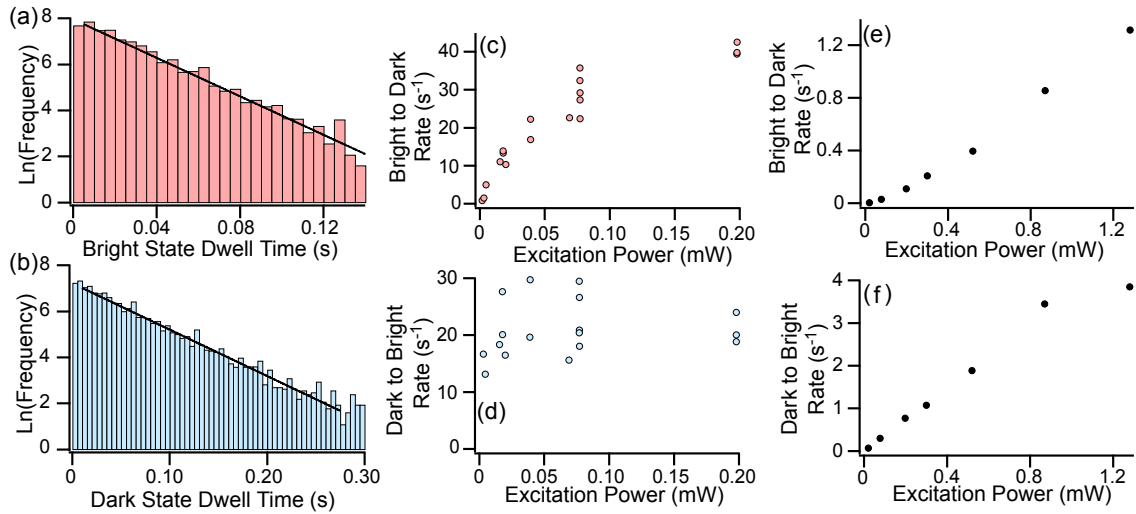

(a-b) Histogram of dwell times in the bright state (a) and dark state (b) for defect 5. Note that each distribution is purely exponential for short times and that $\langle \tau_{BS/DS} \rangle$ is on the order of milliseconds. Similar histograms were acquired for a range of laser powers and the mean transition rate $\langle \tau_{BS/DS} \rangle^{-1}$ was determined for each laser power. The results are presented for defect 5 for the bright state (c) and dark state (d). An identical analysis was performed for defect 3 and the results for the bright state and dark state are shown in (e) and (f), respectively. The dark state dwell times for defect 5 and 3 [shown in (d) and (f), respectively] exhibit qualitatively different trends that strongly suggest a different blinking mechanism is dominant for each defect.

**Figure 7:**

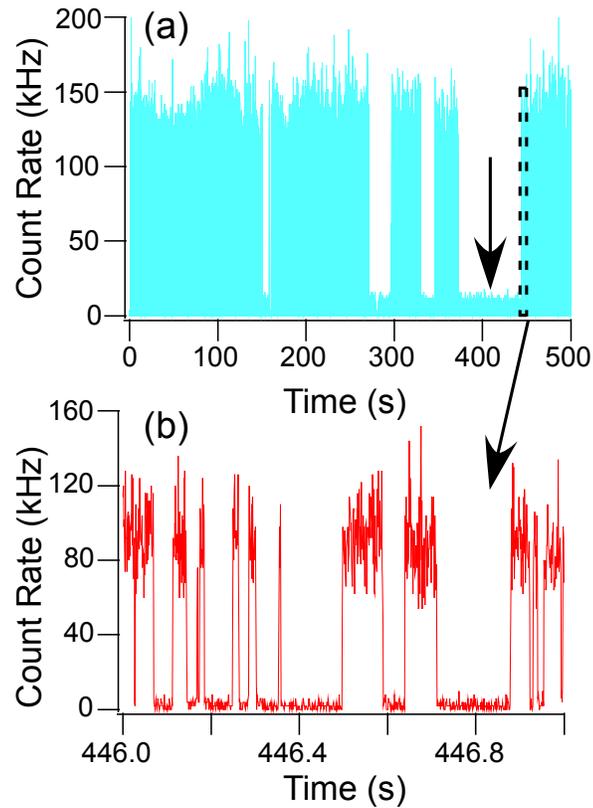

(a) The fluorescence intensity trajectory of defect 5 over a time scale of 500 s revealing anomalously long-lived dark state dwell times. The ordinary blinking of this defect is on millisecond time scales [Figs. 6 (a)-(d)] and is illustrated in (b), which plots a sub-second time window of the fluorescence from (a). In (a) several statistically improbable anomalous dark state dwell times are evident, one of which is denoted by the solid arrow at ~ 400 s. These long-lived dark state dwell times cannot result from the same blinking mechanism that produced the exponential distribution seen in Fig. 6(d) and their existence strongly suggests multiple dark state varieties may jointly couple to a particular defect.